\documentclass[aps,prl,groupedaddress,twocolumn,showpacs]{revtex4-1}
\usepackage[dvips]{graphicx}
\usepackage{dcolumn}
\usepackage{bm}
\usepackage{amssymb}
\usepackage{epstopdf}
\usepackage{amsmath}
\usepackage{color}
\usepackage[dvipdfmx]{hyperref}
\hypersetup{
 setpagesize = false,
 colorlinks = true,
 citecolor = blue,
 linkcolor = blue,
 urlcolor = blue,
 }
\begin{document}

\title{High-temperature ferromagnetism in two-dimensional material MnSn originated from interlayer coupling}

\author{Panjun Feng$^{1}$}\email{These authors contributed equally to this work.}
\author{Xiaohui Zhang$^{2}$}\email{These authors contributed equally to this work.}
\author{Dapeng Liu$^{1}$}
\author{Shuo Zhang$^{1}$}
\author{Xun-Wang Yan$^{1}$}\email{Corresponding author: yanxunwang@163.com}
\author{Z. Y. Xie$^{2}$}\email{Corresponding author: qingtaoxie@ruc.edu.cn}
\date{\today}
\affiliation{$^{1}$College of Physics and Engineering, Qufu Normal University, Qufu, Shandong 273165, China}
\affiliation{$^{2}$Department of Physics, Renmin University of China, Beijing 100872, China.}
\begin{abstract}
 MnSn monolayer synthesized recently is a novel two-dimensional ferromagnetic material with a hexagonal lattice, in which three Mn atom come together to form a trimer, making it remarkably different from other magnetic two-dimensional materials. Most impressively, there happens a sharp increase of  Curie temperature from 54 K to 225 K when the number of layers increase from 1 to 3. However, no quantitative explanation is reported in previous studies. Herein, by means of first-principle calculations method and Monte carlo method, we demonstrate that strong interlayer ferromagnetic coupling is the essential role in enhancing its critical temperature, which act as a magnetic field to stabilize the ferromagnetism in the MnSn multilayers.

\end{abstract}


\maketitle
The magnetic properties of two-dimensional materials have always been a hot topic in condensed matter physics and materials science fields.
Graphene is the first two-dimensional material and possesses excellent mechanical properties, thermal conductivity, and electrical conductivity.
However, non-magnetism prevent graphene from the application in spintronics area. So, various means are used to realize the magnetism in graphene-related systems such as doping, chemical modification, defect and boundary engineering. Unfortunately, such magnetic properties are disordered, non-intrinsic and difficult to control. Until 2016, Je-geun Park proposed the concept of two-dimensional magnetic van der Waals materials\cite{Park2016}. In the same year, two-dimensional antiferromagnetic van der Waals materials NiPS$_3$ and FePS$_3$ were synthesized
\cite{Lee2016,Kuo2016}. In 2017, two-dimensional ferromagnetic materials Cr$_2$Ge$_2$Te$_6$ and CrI$_3$ were discovered for the first time \cite{Gong2017,Huang2017}.  In 2018, room or near room temperature ferromagnetism of transition metal dihalides VSe$_2$ and MnSe$_2$ was discovered \cite{Bonilla2018,OHara2018}.  Since then, the experimental and theoretical study of two-dimensional materials with intrinsic ferromagnetism has become a booming research field.

For the ferromagnetic two-dimensional materials synthesized experimentally, there are three main kinds of structures\cite{Gong2019,Zhang2021a,Hossain2022,Blei2021}. The first one is XY$_3$ structure represented by CrI$_3$, including CrBr$_3$, CrCl$_3$, VI$_3$, VBr$_3$, FePS$_3$, MnPS$_3$, NiPS$_3$, Cr$_2$Ge$_2$Te$_6$, Cr$_2$Si$_2$Te$_6$ and CrCuP$_2$S$_6$. In these two-dimensional magnets, the core structure is XY$_3$ octahedron
 in which Y and X are at the apex and center of the octahedron respectively, and each octahedron shares edges with the three surrounding octahedron. The second typical structure is XY$_2$ structure represented by VSe$_2$, including VS$_2$, MnSe$_2$ and NiI$_2$. Its core structural unit is still an octahedron with a vertex of Y, but each octahedron shares edges with six surrounding octahedrons. The third structure is XY structure, including FeTe and MnSe, in which Y atoms form a tetrahedron and X is in the center of the tetrahedron. So, from the perspective of structure, the two-dimensional magnets synthesized in the experiment are mainly composed of octahedrons or tetrahedrons.
Recently, ferromagnetic MnSn monolayer and multilayer \cite{Yuan2020}, with the distinct structural features from the above materials, were fabricated by molecular beam epitaxy growth method.

\begin{figure}
\begin{center}
\includegraphics[width=7.50cm]{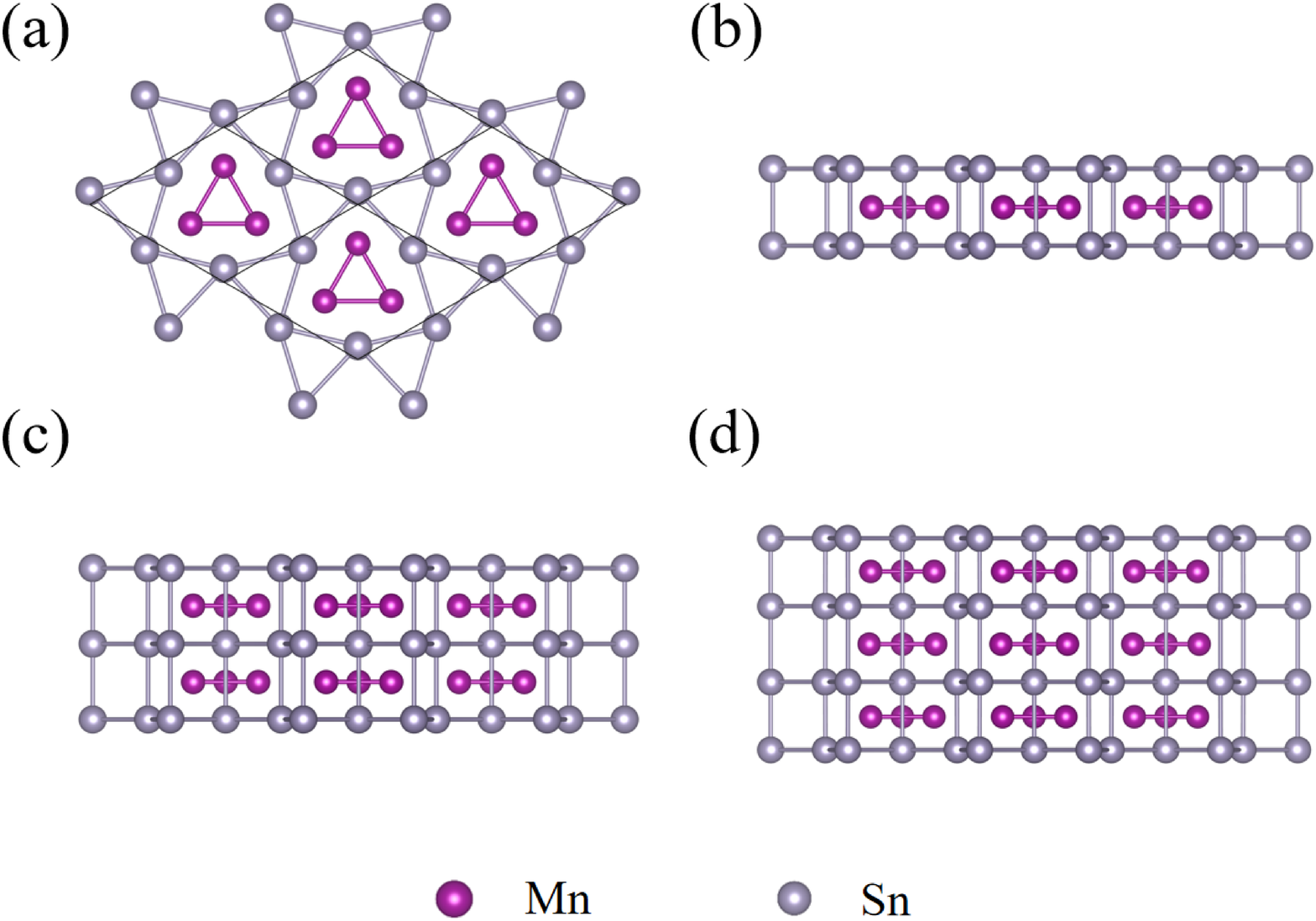}
\caption{The 2 $\times$ 2 unit cell of MnSn monolayer, bilayer, and trilayer. (a) top view, the unit cell is marked with a solid line rhombus; (b) side view of monolayer; (c) side view of bilayer; (d) side view of trilayer.
 } \label{struct-a}
\end{center}
\end{figure}

MnSn monolayer has a hexagonal lattice with the symmetry of $P$-$62m$ space group (No.189), and the lattice parameters are $a$ = $b$ = 6.57 ~\AA. Each unit cell is composed of three Mn atoms and three Sn atoms. More interestingly, three Mn atoms are gathered together to form a trimer with an equilateral triangular shape, and Sn atoms make up a distorted kagome lattice. Atomic structure of MnSn monolayer is shown in Fig. \ref{struct-a}, where (a) and (b) are top view and side view. The Mn trimer and the distorted kagome lattice of Sn atoms are so unique that they do not appear in reported crystal structures.
Based on the common sense that atomic structure determine physical properties, we can deduce that MnSn monolayer would have exotic electronic and magnetic properties.
Most impressively, there exists a sharp increase of Curie temperature from 54 K to 225 K when the number of layers increase from 1 to 4, and until now no quantitative explanation is reported.

In this letter, by means of the first-principles calculation and tensor renormalization group method, we not only explore the unusual electronic and magnetic features of MnSn monolayer  and multilayer, but also demonstrate that the interlayer ferromagnetic coupling play a significant role in realizing the high-temperature ferromagnetism in MnSn multilayer.
In addition, we consider the interlayer magnetic coupling as a magnetic field normal to the MnSn layer, and enclose it into two-dimensional Heisenberg model to evaluate the critical temperature, which provides a practical approach to estimate the critical temperature for multiple layers of magnetic material.

The calculations are carried out based on the plane wave pseudopotential method enclosed in VASP package and the projector augmented-wave (PAW) pseudopotential with Perdew-Burke-Ernzerhof (PBE) functional \cite{PhysRevB.54.11169,PhysRevB.47.558, PhysRevLett.77.3865,PhysRevB.50.17953} is used.
GGA + U method is employed in our calculations because MnSn is $3d$ transition metal compound. The Hubbard U value of 5.66 eV is determined by linear response approach developed by Matteo Cococcioni \cite{Cococcioni2005}.
The plane wave basis cutoff is 600 eV and the convergence thresholds of total energy and force convergence are 10$^{-5}$ eV and 0.01 eV/\AA, respectively.
The interlayer distance was set to 18 \AA~ and a mesh of $24\times 24\times 1$ k-points is used for the Brillouin zone integration.
The temperature of phase transition is evaluated by Monte Carlo method and the $60 \times 60$ lattice is used to represent the magnetic triangle lattice of MnSn layer.

\begin{figure}
\begin{center}
\includegraphics[width=7.50cm]{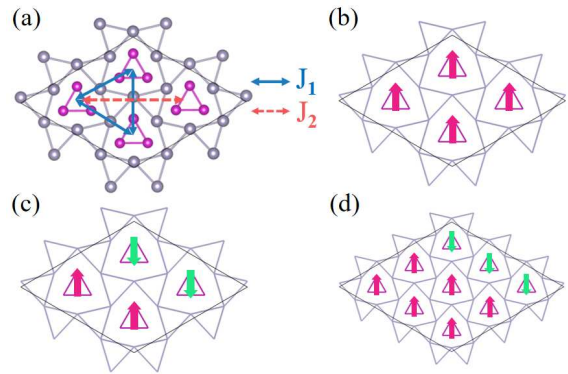}
\caption{ (a) Top view of MnSn monolayer. J$_1$ and J$_2$ are the nearest and next-nearest neighboring magnetic interactions in triangle lattice. (b) FM order; (c) AFM1 ordering; (d) AFM2 ordering. The solid line rhombus represents the magnetic unit cell. To exhibit the magnetic order more clearly, the atomic structure is displayed with the wire frame.
 } \label{order}
\end{center}
\end{figure}

\begin{figure}
\begin{center}
\includegraphics[width=7.50cm]{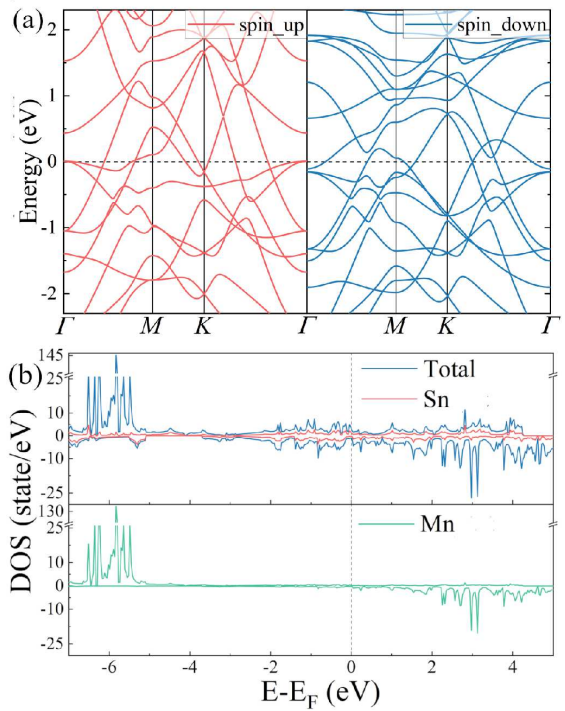}
\caption{Electronic structure of MnSn monolayer. (a) Left panel, spin-up electronic bands; Right panel, spin-down electronic bands. (b) Total density of states and projected density of states on Mn and Sn atoms.
 } \label{band-dos}
\end{center}
\end{figure}

Fig. \ref{struct-a} displays the 2 $\times$ 2 unit cell of MnSn structure, and the top view and side views for monolayer, bilayer, and trilayer are plotted in Fig. \ref{struct-a}(a), (b), (c), and (d).
In the MnSn layer, each Mn trimer acts as one lattice site and all sites make up a triangle lattice.
We first investigate the magnetism of MnSn monolayer.
To find out the magnetic ground state, we calculate the energies of MnSn monolayer with four magnetic orderings revealed by Fig. \ref{order}.
The energies are -39.026 eV, -39.009 eV, and -39.016 eV per unit cell for FM, AFM1, and AFM2 orderings, respectively, indicating that MnSn monolayer on the Sn buffer layer exhibits ferromagnetic behavior.
The band structure of MnSn monolayer in ferromagnetic ordering is presented in Fig. \ref{band-dos}(a), in which multiple bands cross the Fermi level, indicating its metallic behavior. Fig. \ref{band-dos}(b) exhibits the total density of states and projected density of states on Sn and Mn atoms.
The electronic states near Fermi level are dominated by Sn atoms, while the spin polarization of Mn electrons make a decisive contribution to the magnetization of MnSn layer and the moment is mainly located around the Mn atoms.

\begin{figure}
\begin{center}
\includegraphics[width=7.50cm]{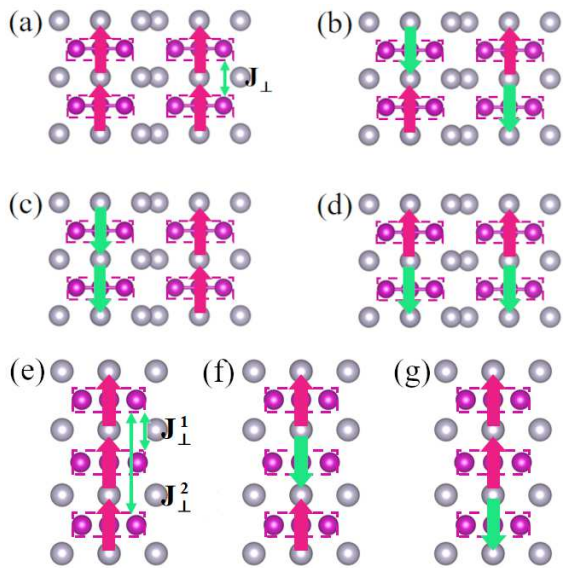}
\caption{Side views of 2 $\times$ 1 $\times$ 2 supercell and 1 $\times$ 1 $\times$ 3 supercell. Red and green arrows indicate the directions of the magnetic moments on Mn trimer. $J_{\bot}$, $J^1_{\bot}$, and $J^2_{\bot}$ are the interlayer magnetic couplings between two nearest neighboring layers and between two next-nearest neighboring layers.
 } \label{InterlayerFM}
\end{center}
\end{figure}

In order to judge whether the ferromagnetic coupling of two adjacent MnSn layers is favorable or not, we calculate the energies of MnSn bilayer and trilayer with different magnetic orders to estimate the interlayer magnetic interations.
For MnSn bilayer, we choose a 1 $\times$ 2 $\times$ 2 supercell and display four magnetic orders from its side view, which are shown in Fig. \ref{InterlayerFM} (a), (b), (c), and (d), corresponding to FM, AFM1, AFM2, and AFM3, respectively.
For AFM1 and AFM2 orderings with the antiparallel moments in one layer, the energy of AFM2 is lower than that of AFM1, indicating that ferromagnetic interlayer coupling is preferred in the case of intralayer antiferromagnetism.
For FM and AFM3 orderings with parallel moments in one layer, the energy of FM is lower than that of AFM3, indicating that ferromagnetic interlayer coupling is also preferred in the case of intralayer ferromagnetism.
So, for MnSn bilayer the interlayer coupling is ferromagnetic.
What about three layers? A 1 $\times$ 1 $\times$ 3 supercell composed of three MnSn layers is built to exhibit different interlayer magnetic couplings, presented in Fig. \ref{InterlayerFM} (f), (g), and (h), reltated to FM$^\prime$, AFM1$^\prime$, and AFM2$^\prime$. From AFM1$^\prime$, AFM2$^\prime$ to FM$^\prime$, energy goes down because of more ferromagnetic couplings.
Therefore, this further demonstrate that the interlayer magnetic coupling is ferromagnetic for MnSn multilayers.

\begin{figure}[htbp]
\begin{center}
\includegraphics[width=8.0cm]{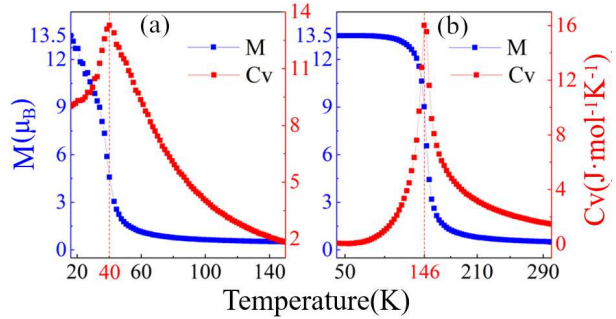}
\caption{The specific heat $C_v$ and average magnetic moment $M$ as functions of temperature for Heisenberg model (a) and Ising model (b) on triangle lattice for MnSn bilayer.
  } \label{Tc}
\end{center}
\end{figure}

Curie temperature is a vital parameter to ferromagnetic materials. At present, Curie temperature lower than room temperature is the main obstacle to limit the applications of two-dimensional ferromagnetic materials in microscopic electronic devices. In our previous work \cite{Feng2022}, we suggest that the dimerization of magnetic ions is a effective approach to realize high-temperature ferromagnetism. In MnSn layer, three Mn ions form a trimer and this lead to large moment on one lattice site. Compared with single Mn ion, Mn trimer can obviously increase the magnetic interaction energy between two lattice sites.
The critical temperature of magnetic phase transition is usually evaluated through solving Heisenberg model with Monte Carlo method \cite{Torelli2018}.
For two-dimensional Heisenberg model, the Hamiltonian is defined as
\begin{equation}
\label{Heisenberg}
H = \sum_{<ij\alpha>}J_{1\alpha}{S}_{i\alpha}{S}_{j\alpha} + \sum_{\ll ij^{\prime}\alpha \gg}J_{2\alpha}{S}_{i\alpha}{S}_{j^{\prime}\alpha} + A\sum_{i}(S_{iz})^2,
\end{equation}
in which the symbol $j$ and $j^\prime$ represent the nearest and next-nearest neighboring sites of $i$ site in triangle lattice, and $\alpha$ is coordinate component $x$, $y$, or $z$. $A$ is the single-site magnetic anisotropic energy.
We perform the non-collinear magnetism calculations with the spin along (1 0 0), (0 1 0), and (0 0 1) axis for different magnetic orderings, and the magnetic exchange interactions can be derived from these energies. There is no significant difference in the exchange interactions along different directions. For MnSn monolayer, the exchange couplings are $J_{1x}$ = $J_{1y}$ = $J_{1z}$ = -5.46 meV/S$^2$ and $J_{2x}$ = $J_{2y}$ = $J_{2z}$ = -1.34 meV/S$^2$, and the single-site magnetic anisotropic energy is -0.636 meV/S$^2$ with easy axis normal to MnSn layer.
The Curie temperature is 40 K, as reflected by the variation of magnetic moment and the specific heat with temperature in Fig. \ref{Tc}(a).

As for MnSn bilayer or multilayer, how to evaluate its Curie temperature from the view of electronic structure simulations is an interesting question.
Hamiltonian of two-dimensional Heisenberg model with an external magnetic field $\vec{h}$ can be written as
\begin{equation}
\label{Heisenberg-2}
H = \sum_{i \neq j}(J_{ij} \vec{S_i} \cdot \vec{S_j}) + \sum_{i}(\vec{h} \cdot \vec{S_i}).
\end{equation}
The interlayer coupling of MnSn bilayer or multilayer is ferromagnetic coupling, as demonstrated above.
For a selected layer, the ferromagnetic interaction from its adjacent layer plays a similar role of an external field normal to MnSn layer.
Hence, it is possible that we choose one MnSn layer to compute the Curie temperature based on Heisenberg model.
\begin{equation}
\begin{aligned}
\label{Heisenberg}
H = \sum_{<ij\alpha>}J_{1\alpha}{S}_{i\alpha}{S}_{j\alpha} + \sum_{\ll ij^{\prime}\alpha \gg}J_{2\alpha}{S}_{i\alpha}{S}_{j^{\prime}\alpha} \\
 + A\sum_{i}(S_{iz})^2 + \sum_{i}J_{\bot}(S_{\bot z}S_{iz}),
\end{aligned}
\end{equation}
where the last term is the interlayer coupling.
Apart from the last term, the Hamiltonian in Expression (3) and (1) are the same.
For MnSn bilayer, the interlayer coupling $J_{\bot}$ is 84.16 meV/S$^2$. 
Compared to the exchange couplings of $J_{1x}$ = $J_{1y}$ = $J_{1z}$ = -5.46 meV/S$^2$, the interlayer coupling $J_{\bot}$ of -84.16 meV/S$^2$ is large value. It acts as a strong magnetic field on MnSn layer, which forces the moments in a certain direction and leads to strong anisotropy.
In this case, the magnetic interactions in MnSn layer can be described by Ising model. The Curie temperature of 146 K is determined by the following Ising model,
\begin{equation}
\label{Ising}
H = \sum_{<ij>}J_{1}{S}_{i}{S}_{j} + \sum_{\ll ij^{\prime} \gg}J_{2}{S}_{i}{S}_{j^{\prime}},
\end{equation}
where $J_1$ = -5.46 mev/S$^2$ and $J_2$ = -1.34 mev/S$^2$.
The curve of magnetic phase transitions for MnSn bilayer is shown in Fig. \ref{Tc}(b). The Curie temperatures of MnSn bilayer is 146 K, much higher than the value of 40 K for monolayer, which is because strong interlayer ferromagnetic interaction stabilize its ferromagnetism.
For MnSn trilayer, the interlayer couplings $J^1_{\bot}$ and $J^2_{\bot}$ are -102.6 meV/S$^2$ and -4.7 meV/S$^2$, and the sum of $J^1_{\bot}$  and $J^2_{\bot}$ is $J_{\bot}$ = -107.3 meV/S$^2$, higher than -84.16 meV/S$^2$ in case of bilayer.
From bilayer to trilayer, the increase of interlayer coupling $J_{\bot}$ is due to the coupling from the third layer, which results in stronger ferromagnetism. This is consistent with the experimental result that from three layers to four layers the Curie temperature of MnSn film increases from 225 K to 235 K.

In summary, we systematically investigate the magnetic properties of MnSn monolayer and the MnSn multilayer by means of the first-principle calculation and Monte Carlo method. From MnSn monolayer to bilayer, the calculated Curie temperature varies from 40 K to 146 K owning to the inclusion of strong ferromagnetic interactions between layers.
So, we demonstrate that strong interlayer ferromagnetic coupling plays an essential role in enhancing its critical temperature, which act as a magnetic field to stabilize the ferromagnetism in the MnSn multilayers.
Our results not only explain the sharp increase of Curie temperature of MnSn film from 54 K to 225 K in experiments but also reveal that enhancing interlayer coupling is a new routine to realize high-temperature ferromagnetism in two-dimensional materials.

This work was supported by the National Natural Science Foundation of China (Grants Nos. 11974207, 11974194, 11774420, 12074040), the National R\&D Program of China (Grants No. 2017YFA0302900), and the Major Basic Program of Natural Science Foundation of Shandong Province (Grant No. ZR2021ZD01).

\bibliography{Ref}


\end{document}